\documentclass{PoS}
\usepackage[mathcal]{eucal}

\title{Dark matter searches with the IceCube Upgrade}
\ShortTitle{Dark matter searches with the IceCube Upgrade}

\author{
The IceCube Collaboration\footnote{For collaboration list, see PoS(ICRC2019) 1177.}\\
{\itshape \href{http://icecube.wisc.edu/collaboration/authors/icrc19_icecube}{http://icecube.wisc.edu/collaboration/authors/icrc19\_icecube}}\\
E-mail: \email{sebastian.baur@icecube.wisc.edu}
}

\abstract{
Weakly Interacting Massive Particles (WIMPs) are well-motivated candidates for Dark Matter (DM). WIMP models often include self-annihilation into Standard Model particles such as neutrinos which could potentially be detected by the IceCube Neutrino Observatory. Various searches for a dark matter induced signal have been performed with the existing IceCube detector. However, since there is so far no evidence for WIMPs at TeV scales, more attention is brought to DM candidates at GeV masses, for which the IceCube detector is not sensitive due to its energy threshold. The IceCube collaboration is currently preparing the construction of the IceCube Upgrade which is planned to be deployed in the 2022/2023 South Pole summer season. The IceCube Upgrade will consist of 7 new in-ice strings with about 700 additional optical sensors. This dense sensor array inside the IceCube-DeepCore volume will enhance the reconstruction capability of few-GeV neutrinos. We present first studies on the potential improvements of this upgrade on IceCube's sensitivity to Dark Matter annihilating in the Galactic Center. 

\vspace{4mm}
{\bfseries Corresponding authors:}
\speaker{Sebastian Baur}$^{1}$\\
{$^{1}$ \itshape Universit\'e Libre de Bruxelles}
}
\FullConference{36th International Cosmic Ray Conference -ICRC2019-\\
		July 24th - August 1st, 2019\\
		Madison, WI, U.S.A.}

\begin{document}

\newcommand{\numu}{$\nu_\mathrm{\mu}$}
\newcommand{\numubar}{$\bar{\nu}_\mathrm{\mu}$}
\newcommand{\sv}{$\langle\sigma v\rangle$}
\newcommand{\mass}{$m_\mathrm{\chi}$}

\section{Introduction}\label{sec:intro}
A large variety of observations, such as the rotational motion of stars in spiral galaxies or the power spectrum of the cosmic microwave background, suggest the existence of invisible non-baryonic dark matter (DM). The contribution of DM to the total energy content of the universe is assumed to be significant in most models. The question of the precise nature of dark matter besides the gravitational effects, however, is yet to be solved. In many scenarios, Weakly Interacting Massive Particles (WIMPs) may create standard-model particles via decay or self-annihilation which can be observed with space- or ground-based experiments, providing indirect evidence for the particle nature of dark matter. Many searches for these signatures have been performed with various experiments without any indication so far.

Neutrinos appear as final state particles in many possible annihilation processes. The IceCube neutrino observatory therefore has an ideal opportunity to study or constrain the properties of DM particles, see ~\cite{Sanchez:2019rra} for a recent review. Since the allowed parameter space for $\mathcal{O}\left(\mathrm{TeV}\right)$ WIMPs is becoming more and more constrained by both direct and indirect measurements, increased attention is brought again to lighter dark matter candidates (see~\cite{Battaglieri:2017aum} for a recent review).

With the foreseen upgrade of IceCube, the telescope's efficiency to detect and reconstruct for low-energy neutrinos is expected to be significantly improved. As a consequence, the sensitivity to the velocity averaged cross section \sv{} of dark matter annihilations in the galaxy will equally improve. In this contribution, annihilations of dark matter in the galactic DM halo with masses between 2 and 150\,GeV are considered and sensitivities to the \sv{} are calculated for the upgraded IceCube telescope. A method very similar to previous and ongoing analyses of the IceCube collaboration is chosen in order to demonstrate the impact of the upgrade on the detector performance.

\section{The upgrade to the IceCube neutrino observatory}\label{sec:icecube}
\begin{figure}
    \centering
    \includegraphics[width=0.5\linewidth]{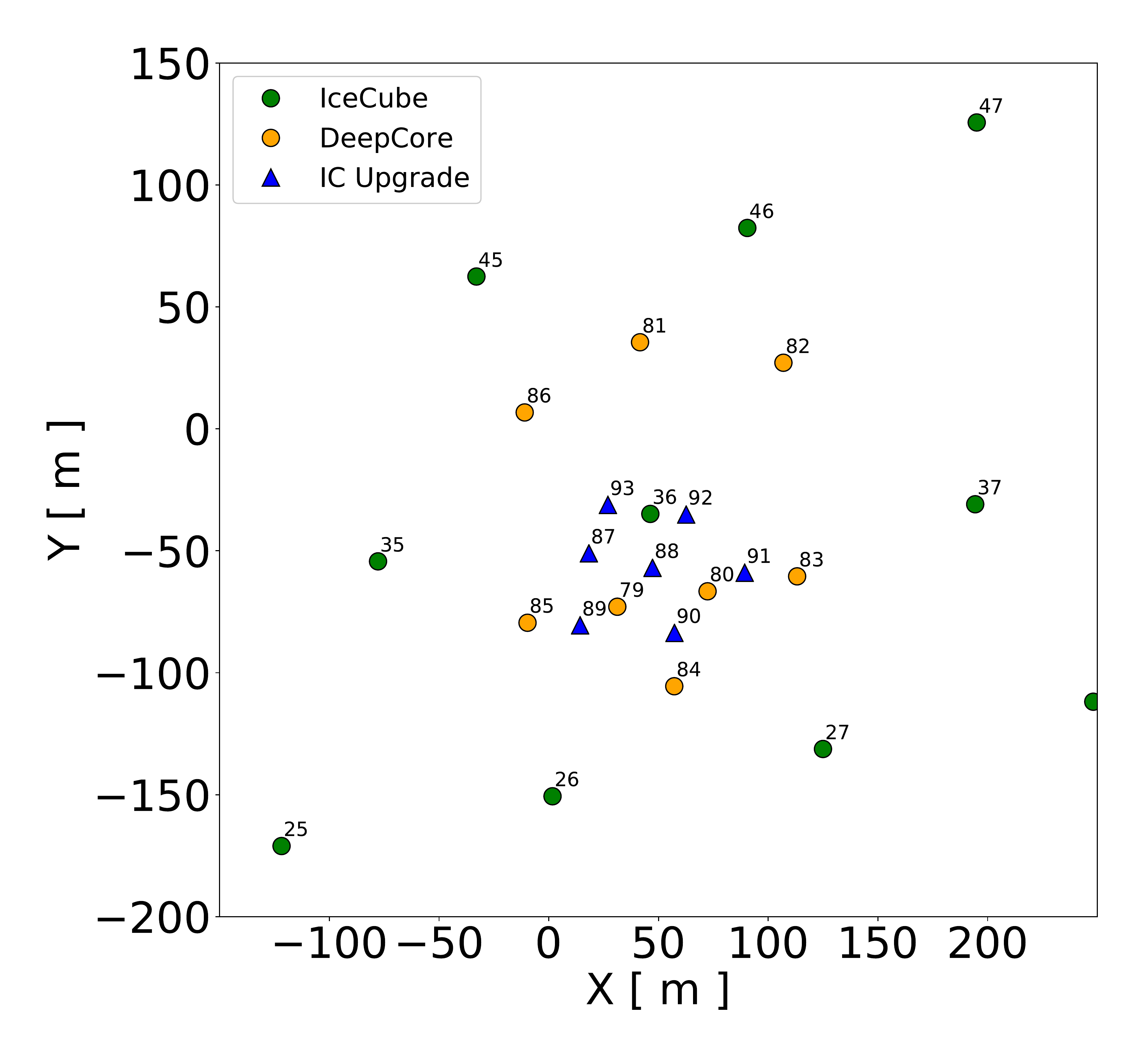}\hfill
    \includegraphics[width=0.4\linewidth]{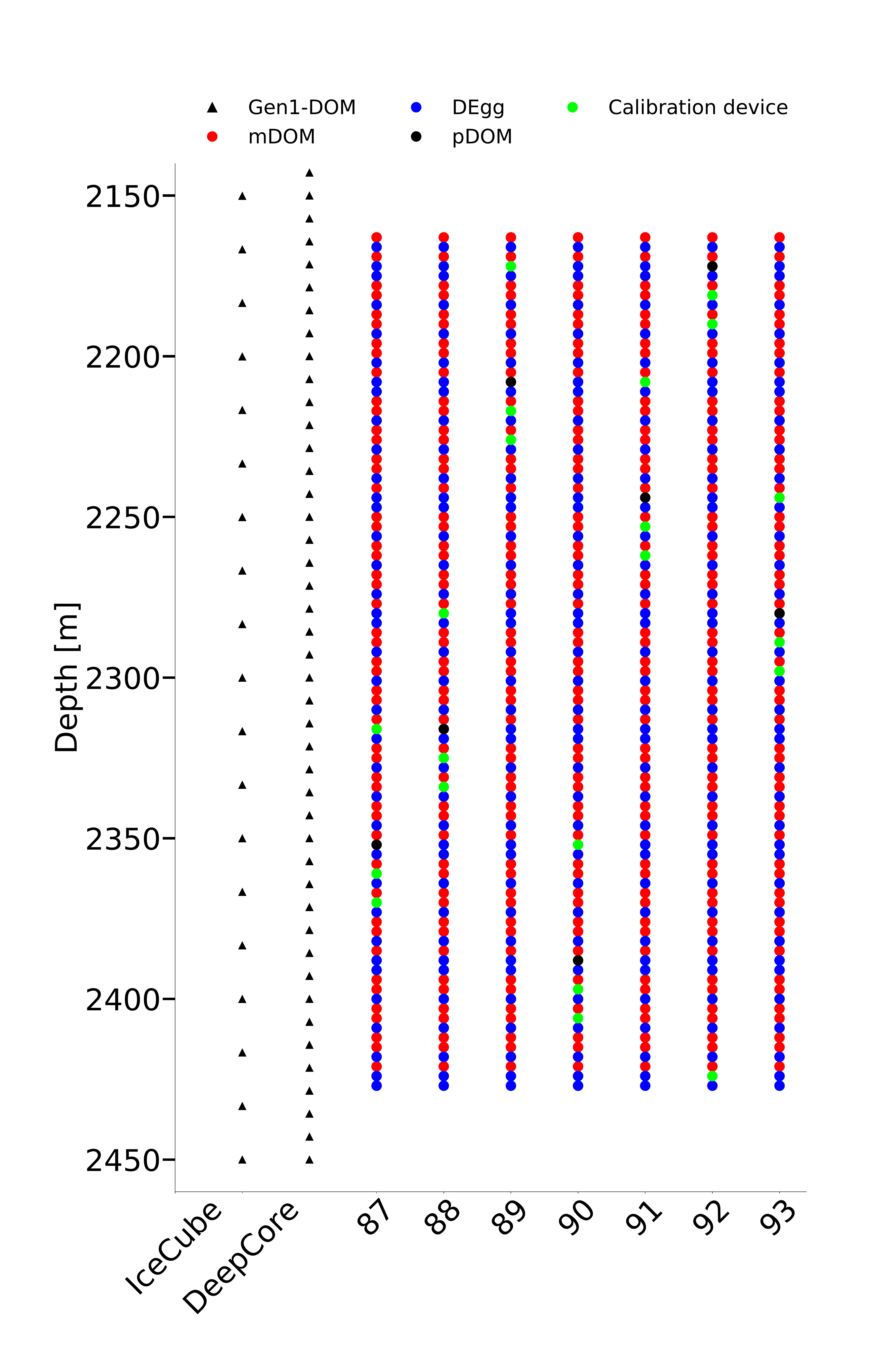}\hfill
    \caption{Illustrations of the planned layout of the IceCube Upgrade compared to the existing IceCube instrumentation. The string position with respect to the existing strings is shown in the left, the density of optical modules along the strings in the right figure.}
    \label{fig:upgradeGeometry}
\end{figure}

IceCube is a cubic-kilometer neutrino detector installed in the ice at the geographic South Pole~\cite{Aartsen:2016nxy} between depths of 1450 m and 2450 m. Reconstruction of the direction, energy and flavor of neutrinos relies on the optical detection of Cherenkov radiation emitted by charged particles produced in the interactions of neutrinos in the surrounding ice or the nearby bedrock. The DeepCore subarray includes 8 densely instrumented strings optimized for low energies. IceCube is currently preparing for an upgrade to be deployed during the 2022/2023 South Pole summer season. This upgrade will consist of 7 new in-ice strings with about 700 additional optical sensors. The new strings will be located inside the existing DeepCore array. The vertical spacing between individual sensors will be reduced to about 3\,m, compared to 7\,m for the DeepCore strings. This denser sensor array is expected to improve the detection efficiency and reconstruction capability of neutrinos with energies down to 1\,GeV.  The foreseen position of these new strings as used in this study is shown in Fig.~\ref{fig:upgradeGeometry}, and a more detailed description of the design and science goals is given in~\cite{Ishihara:2019icrc}.

\begin{figure}
    \centering
    \includegraphics[width=0.7\linewidth]{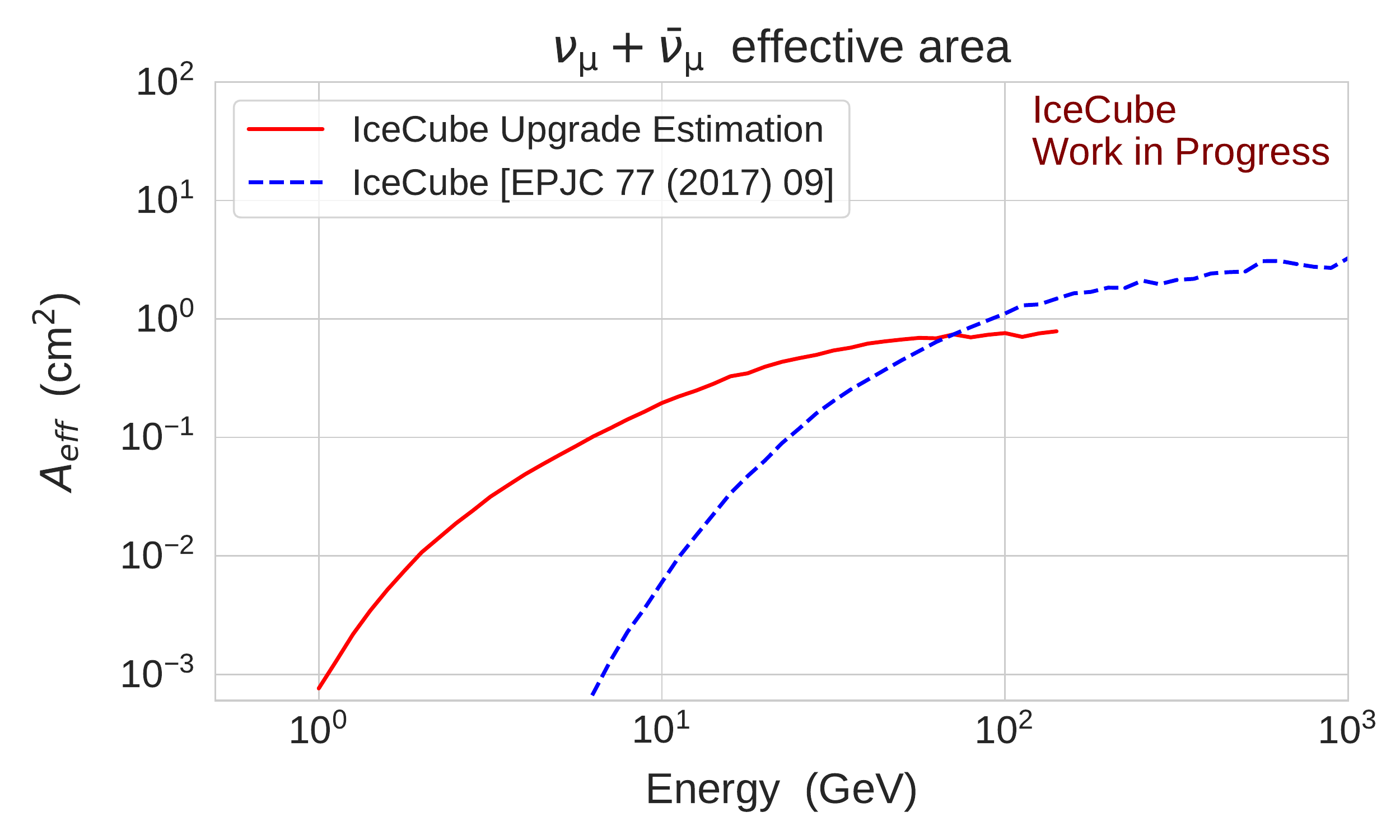}
    \caption{Effective area for the detection of \numu and \numubar. The estimation for the IceCube Upgrade is compared to the effective area of a previous analysis performed by IceCube~\cite{Aartsen:2017ulx}.}
    \label{fig:effectiveArea}
\end{figure}

A preliminary event selection is developed based on simulation of the upgraded detector with the aim to enhance the acceptance to low-energy neutrinos. In contrast to previous analyses, the event selection is not tuned to a specific signal distribution caused by dark matter annihilation. Instead, a maximal rate of atmospheric neutrinos with interaction vertices inside or close to the upgraded volume is desired while rejecting a large fraction of atmospheric muons. The distribution in space and time of the pulses recorded in the array of photo-multipliers allows to efficiently discriminate through-going muons from events starting inside or close to the upgraded volume. This significantly reduces the amount of atmospheric muons passing the selection. The selection is the same for both up- and down-going events\footnote{The acceptance to up-going neutrinos could be improved by taking directional information into account. This would, however, not impact this study since the Galactic Center is, for IceCube, always seen above the horizon.}. Once neutrino energies become larger than about 80\,GeV, the resulting signal patterns are elongated enough so that they are rejected by this selection. It is therefore found that the considered selection is only effective for primary neutrinos with an energy smaller than 100\,GeV. Furthermore, the detector is assumed to be inefficient for neutrino energies below 1\,GeV -- more detailed studies will be necessary to ensure that sub-GeV events can be reliably selected and reconstructed. The final event rates after the event selection are estimated to be 2.68 (1.07)\,mHz for atmospheric muon (electron) neutrinos and 0.37\,mHz for atmospheric muons. The acceptance of this event selection to neutrinos is shown in Fig.~\ref{fig:effectiveArea} in terms of the telescope's effective area for \numu and \numubar as a function of the primary neutrino energy. The detailed event-by-event reconstruction with the upgraded detector is still under development. For the present study, a parameterized reconstruction is therefore applied. This parameterization is based on results obtained with reconstructed $\nu_\mathrm{e}$ events, whose angular resolution is typically worse than for \numu{} events. Nevertheless, the obtained resolutions are improved compared to the current IceCube detector due to the smaller spacing between sensors as well as the multi-PMT design of the individual sensors~\cite{Classen:2019icrc}. All parameterizations used in this analysis can therefore be considered conservative.

\section{Sensitivity calculation to dark matter signals from the Galactic Center}\label{sec:analysis}
The flux of final state neutrinos arriving at Earth for given dark matter particles with mass \mass{} and a velocity-averaged annihilation cross section \sv{} is given by
\begin{equation}
\label{eq:flux}
\frac{\mathrm{d}\Phi_\mathrm{\nu}}{\mathrm{d}E_\mathrm{\nu}} = \frac{1}{4\pi m_\mathrm{\chi}^2}\frac{\langle\sigma v\rangle}{2}\frac{\mathrm{d}N}{\mathrm{d}E_\mathrm{\nu}} \mathcal{J}_{\mathrm{\psi}}~\mathrm{,}
\end{equation}
where $\mathrm{d}N/\mathrm{d}E_\mathrm{\nu}$ is the energy spectrum of the final state neutrinos and $\mathcal{J}_{\mathrm{\psi}}$ the integral of the squared DM density along the line of sight in a given direction $\psi$. In this work, the annihilation spectra provided in~\cite{Cirelli:2010xx,Ciafaloni:2010ti} are used, assuming a branching ratio of 100\% into either $\mu^+\mu^-$, $\tau^+\tau^-$, $b\bar{b}$ or \numu\numubar. Furthermore, two different parameterizations of the galactic DM halo are used, namely Navarro-Frenk-White (NFW)~\cite{Navarro:1995iw} and Burkert~\cite{Burkert:1995yz} profiles with parameters fitted from astrophysical observations presented in~\cite{Nesti:2013uwa}. The directional factors $\mathcal{J}_{\mathrm{\psi}}$ of both halo models are calculated using the \textit{clumpy} software tool~\cite{Hutten:2018aix}. Due to the effective rejection of atmospheric muons, the dominant background is the flux of atmospheric neutrinos for which the energy spectrum as well as angular distribution provided in~\cite{Honda:2015fha} are used.

\begin{figure}
    \centering
    \includegraphics[width=0.49\linewidth]{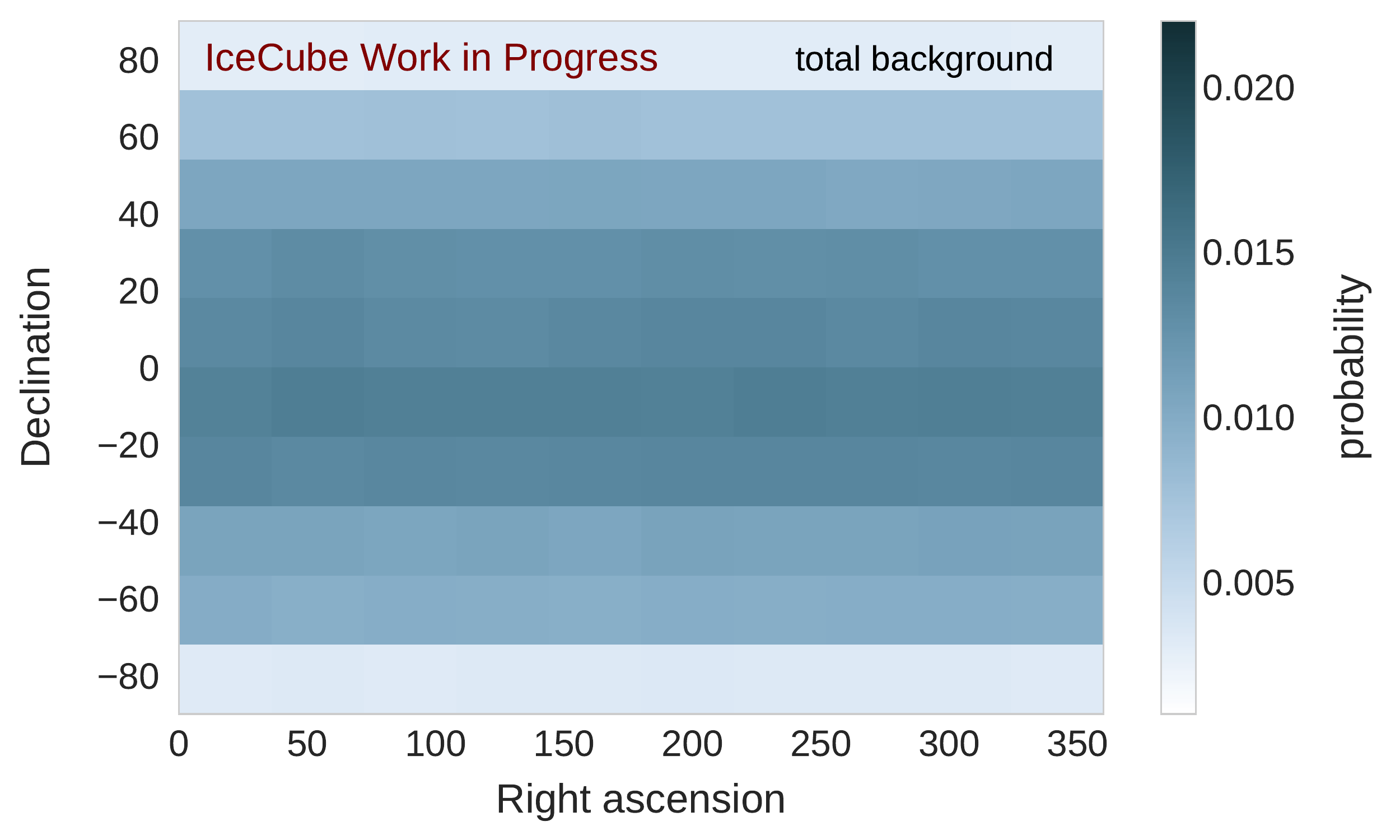}\hfill
    \includegraphics[width=0.49\linewidth]{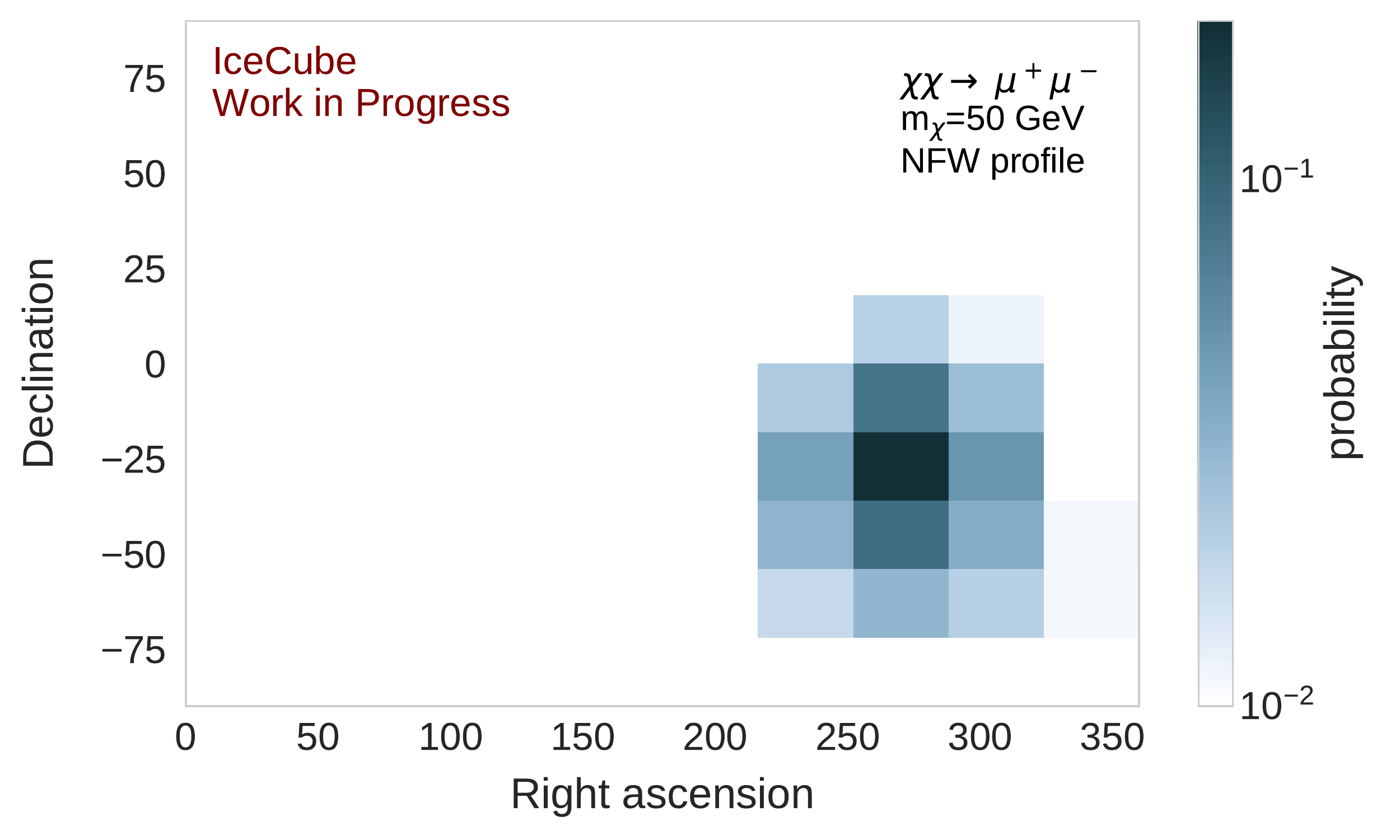}
    \caption{Probability distribution of the expected background due to atmospheric muons and neutrinos (left) and for the expected signal of a DM particle with mass \mass{}=50\,GeV annihilating into $\mu^+\mu^-$ assuming the NFW halo profile (right).}
    \label{fig:PDFs}
\end{figure}

As a result, two-dimensional probability density functions (PDFs) are calculated for both signal and background for a combination of all neutrino flavors. These PDFs are composed of 10 bins in right ascension and 10 bins in declination. This binning scheme is similar to those used in previous and ongoing analyses performed by the IceCube collaboration~\cite{Aartsen:2017ulx,Iovine:2019icrc}. In this way, a direct comparison to previously obtained results is possible and the relative impact of the upgrade is emphasized. The background PDF as well as an example for a signal PDF for a DM particle with mass \mass{}=50\,GeV annihilating into $\mu^+\mu^-$ assuming the NFW halo profile are shown in fig.~\ref{fig:PDFs}. It can be seen that dark matter annihilation results in a strong excess of neutrinos coming from the direction of the galactic center, while the expected background is more uniform and does not favor a specific direction.

In the same way as in~\cite{Aartsen:2017ulx,Iovine:2019icrc}, a likelihood function $\mathcal{L}\left(\mu\right)$ is defined as the product of Poisson probabilities $\mathcal{P}$
\begin{equation}
\label{eq:likelihood}
\mathcal{L}\left(\mu\right) = \prod_{i=1}^{N_{\mathrm{bins}}} \mathcal{P}\left( n_{\mathrm{obs}}^{\mathrm{i}} | n_{\mathrm{obs}}^{\mathrm{tot}} \cdot f^{\mathrm{i}}\left(\mu\right)\right)~\mathrm{,}
\end{equation}
where $n_{\mathrm{obs}}^{\mathrm{i}}$ is the expected number of observed events in bin $i$, $n_{\mathrm{obs}}^{\mathrm{tot}}$ is the total number of observed events, and $f^{\mathrm{i}}\left(\mu\right)$ is the fraction of events expected in bin $i$ assuming a certain signal-to-background ratio $\mu$.

\section{Results and conclusions}\label{sec:results}
\begin{figure}
    \centering
    \includegraphics[width=0.49\linewidth]{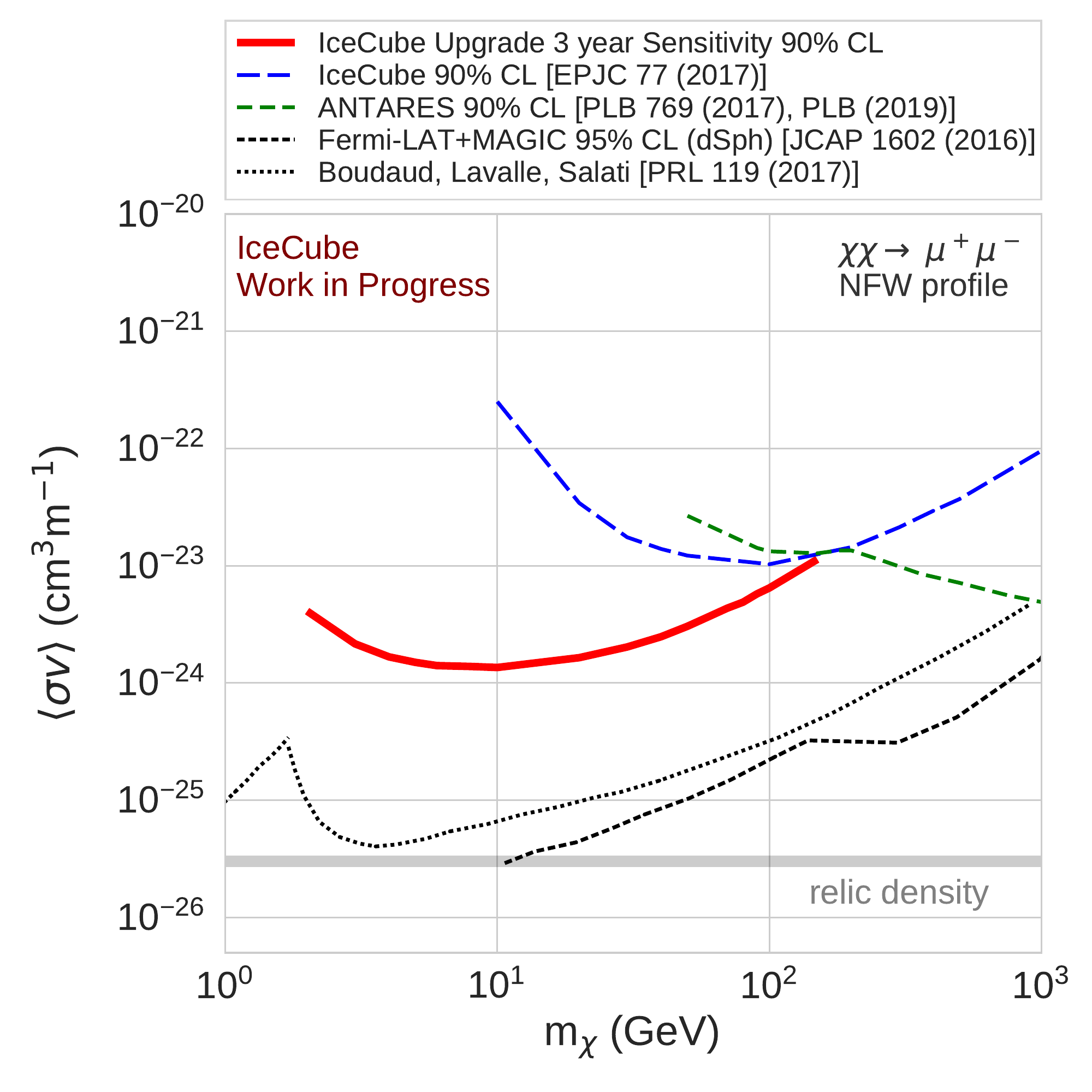}\hfill
    \includegraphics[width=0.49\linewidth]{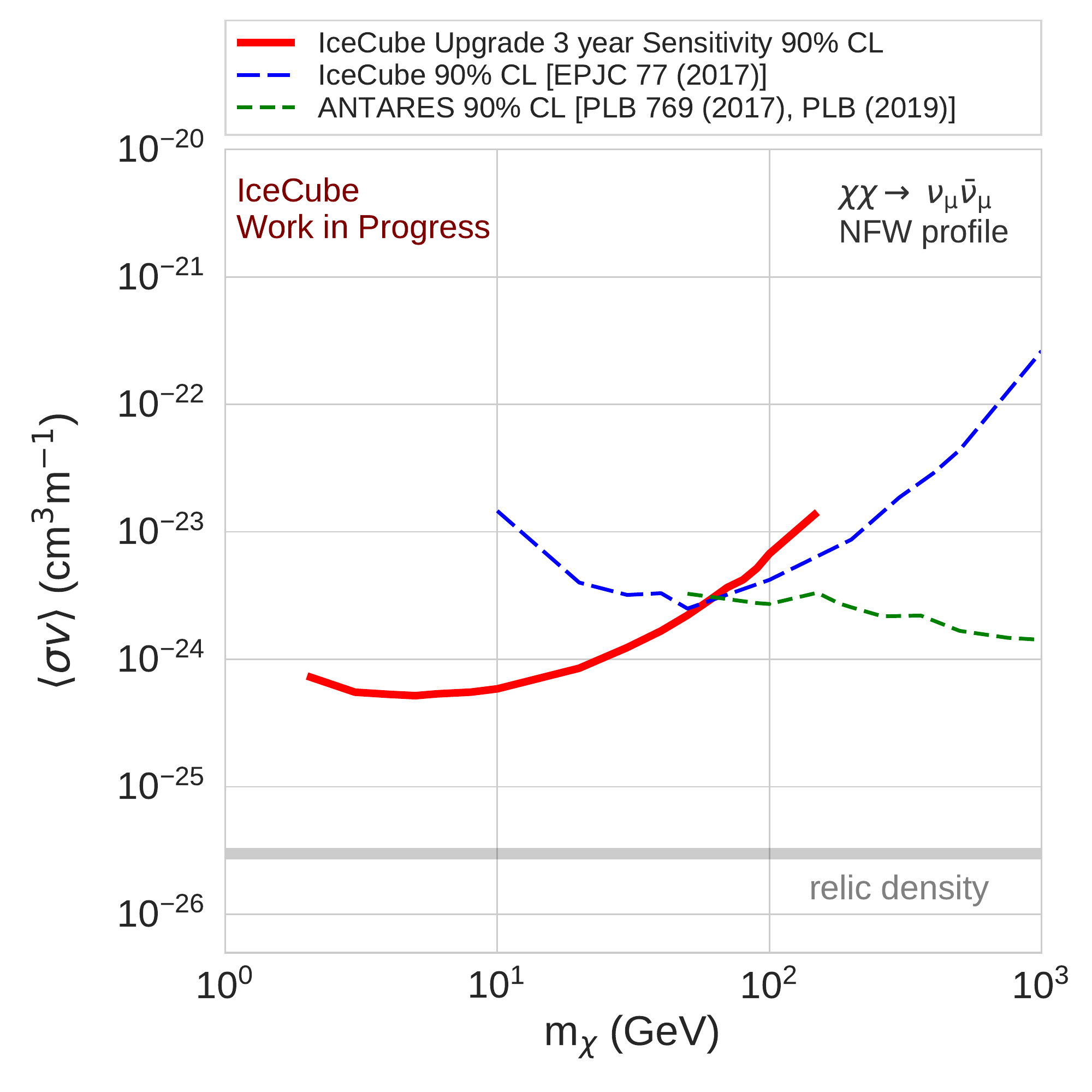}
    \caption{Expected sensitivities on the dark matter self-annihilation cross-section \sv{} with three years of data with the IceCube Upgrade as a function of the dark matter mass \mass. Shown are the annihilation channels into $\mu^+\mu^-$ (left) and \numu\numubar (right) assuming an NFW halo profile. Upper limits obtained by IceCube~\cite{Aartsen:2017ulx} and ANTARES~\cite{Albert:2016emp} as well as the Fermi-LAT and MAGIC~\cite{Ahnen:2016qkx} telescopes and limits using data of the Voyager1 and AMS-02 satellites~\cite{Boudaud:2016mos} are shown for comparison.}
    \label{fig:sensitivities_global}
\end{figure}

\begin{figure}
    \centering
    \includegraphics[width=0.49\linewidth]{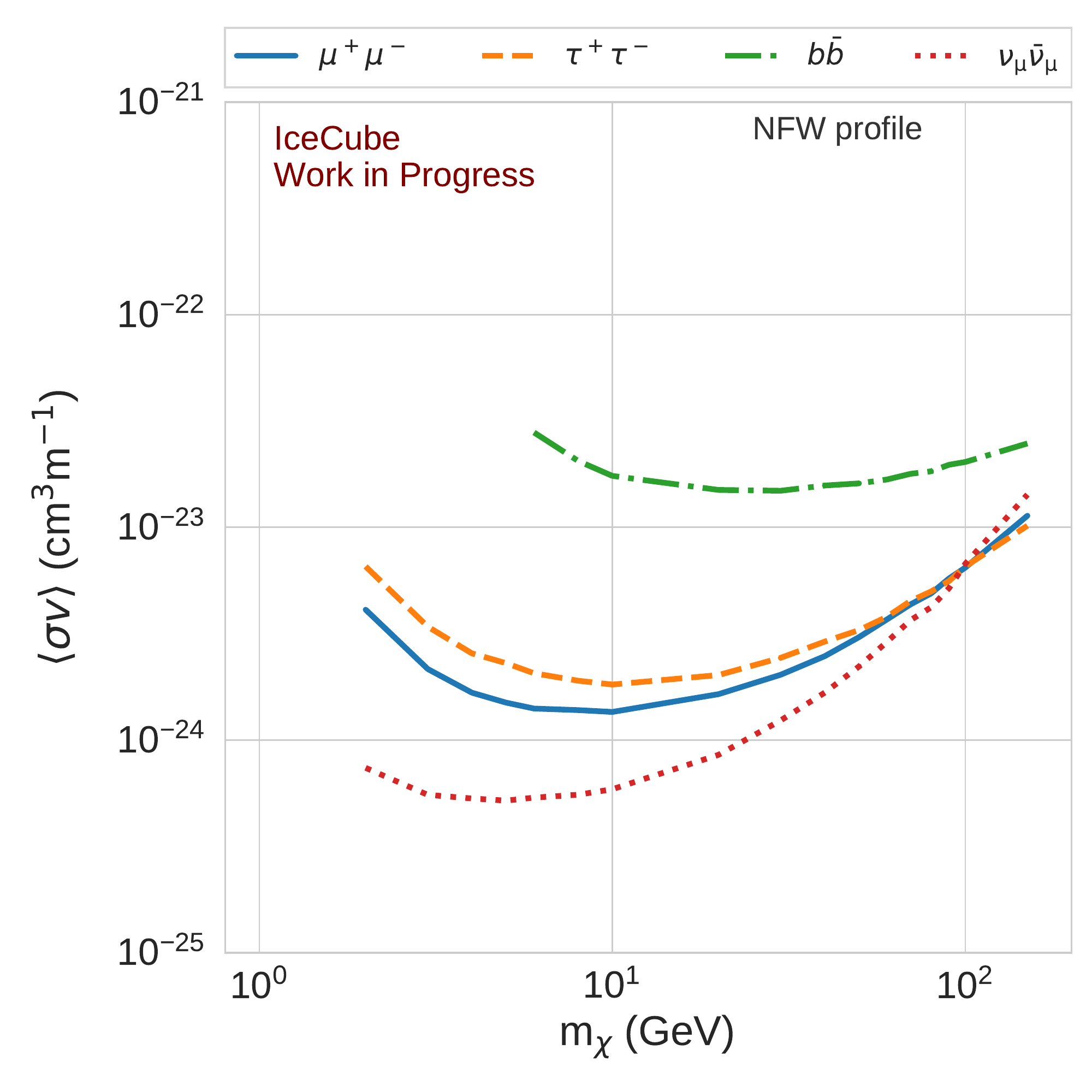}\hfill
    \includegraphics[width=0.49\linewidth]{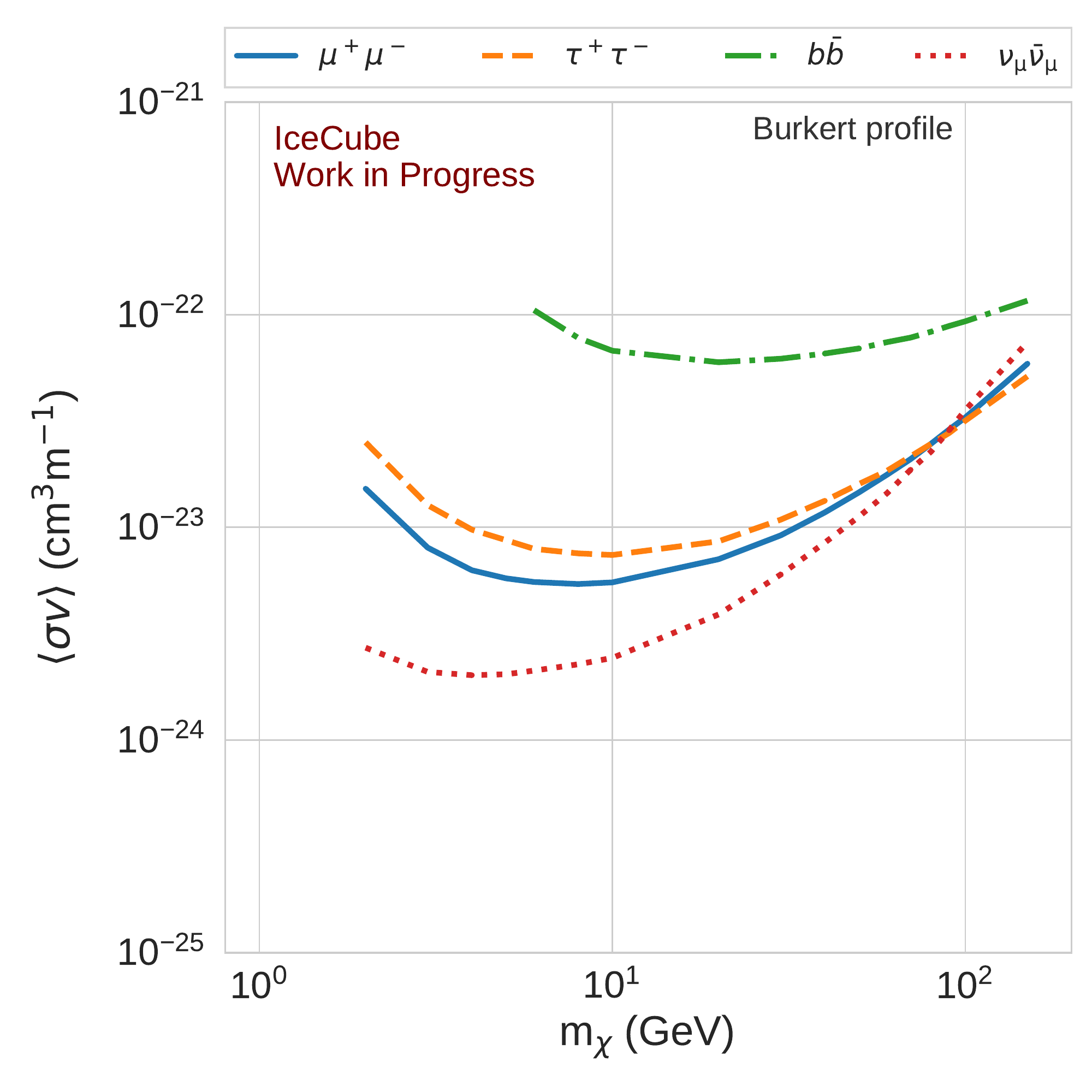}
    \caption{Expected sensitivities on the dark matter self-annihilation cross-section \sv{} with three years of data with the IceCube Upgrade as a function of the dark matter mass \mass. Shown are annihilation into $\mu^+\mu^-$, $\tau^+\tau^-$, $b\bar{b}$, and \numu\numubar assuming a 100\% branching ratio each for the NFW (left) and Burkert (right) halo profiles.}
    \label{fig:sensitivities_ICUonly}
\end{figure}

In order to obtain sensitivities comparable to previous works, three years of data taking are assumed also for the upgraded IceCube detector.
The likelihood $\mathcal{L}$ as defined in Eq.~\ref{eq:likelihood} is evaluated with respect to $\mu$, compared to the background-only hypothesis, and used to calculate the telescope's sensitivity at the 90\% confidence level according to the Feldman-Cousins method~\cite{Feldman:1997qc}. The resulting signal fraction is then converted to sensitivities on \sv{} according to Eq.~\ref{eq:flux}. The estimated sensitivities are shown in Figs.~\ref{fig:sensitivities_global} and \ref{fig:sensitivities_ICUonly}. Upper limits obtained by IceCube~\cite{Aartsen:2017ulx} and ANTARES~\cite{Albert:2016emp} are shown alongside results by the Fermi-LAT and MAGIC~\cite{Ahnen:2016qkx} telescopes (using gamma rays as final state messengers) and a limit derived by Boudaud et al. using $e^\pm$ data by the Voyager1 and AMS-02 satellites~\cite{Boudaud:2016mos}.

With the improved efficiency to detect low-energetic neutrinos and the better angular resolution, the foreseen upgrade of IceCube will significantly improve the detector's abilities to constrain the dark matter self-annihilation cross-section \sv{} for dark matter masses below 100\,GeV. This is especially the case for annihilation channels, where many low-energy neutrinos are created through secondary electro-weak processes. With the upgrade in place, IceCube will for the first time be able to probe DM annihilation for masses well below 10\,GeV. Since the assumptions and parameterizations used are mostly conservative and especially the statistical method is not tuned to the performance and angular resolution of the upgraded detector, this study does not provide the best-possible sensitivity but rather a realistic estimation of the improvement with respect to previous results. Comparable improvements can be expected for dark matter searches using the Sun, the Earth, or other sources. Detailed studies however remain to be performed.

\bibliographystyle{ICRC}
\bibliography{references}

\end{document}